\input harvmac
\input epsf
\baselineskip=.55truecm
\Title{\vbox{\hbox{HUTP--96/A030}\hbox{hep-th/9608186}}}
{\vbox{\centerline{A Comment on BPS States in 
F-theory}
\centerline{in 8 Dimensions}}}
\vskip .1in
\centerline{\sl A. Johansen\foot{E-mail: johansen@string.harvard.edu}}
\vskip .2in
\centerline{\it Lyman Laboratory of Physics, Harvard University}
\centerline{\it Cambridge, MA 02138, USA}
\vskip .2in
\centerline{ABSTRACT}
\vskip .2in

We study some
aspects of enhanced gauge symmetries in F-theory
compactified on K3.
We find open string configurations
connecting various 7-branes which 
represent stable BPS states. 
In this approach we recover $D_n$ and $E_n$ 
gauge groups previously found
from an analysis of sigularities of the moduli space
of elliptically fibered K3 manifolds
as well as 
examples of non-perturbative realizations of $A_n$ groups.

\Date{\bf August 1996}
{\it 1.Introduction}

Recently great progress has been made in our understanding of  
enhanced gauge symmetries and matter in 
compactifications of F-theory
\ref\vafa{C. Vafa, Nucl. Phys. {\bf B 469} (1996) 403, hep-th/9602022.},
\ref\MV{D. Morrison and C. Vafa, `Compactifications of F-theory on 
Calabi-Yau Threefolds-
I, II'', hep-th/9602114, hep-th/9603161.}, 
\ref\kucha{M. Bershadsky, K. Intriligator,
S. Kachru, D. Morrison, V. Sadov, and C. Vafa, `Geometric 
Singularities and Enhanced
Gauge Symmetries', hep-th/9605200.}, \ref\KV{S. Katz and C. Vafa,
`Matter From Geometry', hep-th/9606086.}.
Although the appearance of enhanced gauge symmetries 
and matter 
in compactifications of 
type II \ref\BSVa{M. Bershadsky, V. Sadov and C. Vafa, 
Nucl. Phys. {\bf B 463} (1996) 398, hep-th/9510225.}
\ref\KMP{S. Katz, D.R. Morrison and
M.R. Plesser, `Enhanced Gauge 
Symmetry in Type II String Theory', hep-th/9601108.}
or F-theory 
has been understood 
in a general geometric framework through an
analysis of singularities
it is instructive to try to understand it in terms of 
BPS states produced by open strings with various $(p,q)$ charges \ref\john{J. Schwarz, Phys. Lett. {\bf B 360} 
(1995) 13, hep-th/9508143.} connecting
($SL(2,{\bf Z})$ transformed) Dirichlet branes.
Such an analysis can provide us with an additional evidence for string-string 
duality as well as give us a better understanding of quantum 
field theory applications of the latter
\ref\three{T. Banks, M.R. Douglas and N. Seiberg,
`Probing F-theory with Branes', hep-th/9605199\semi
N. Seiberg, `IR Dynamics on Branes and Space-Time 
Geometry', hep-th/9606017.}.

More evidence for the duality between F-theory on an elliptic K3 
and the heterotic string compactified on a $T^2$ has been found in 
an analysis of the case of K3 corresponding to 
constant background for dilaton-axion fields
\ref\sen{A. Sen, `F-theory and Orbifolds',
hep-th/9605150.}, \ref\khvost{K. Dasgupta and S. Mukhi,
`F-theory at Constant Coupling', hep-th/9606044.}.
In particular it has been shown \sen\ that F-theory on $T^4/{\bf Z}_2$
is equivalent to an orientifold of type IIB on $T^2$, which in turn,
is related by a $T$-duality transformation 
to a type I theory on $T^2$.
In this description an enhanced $SO(8)$ gauge symmetry appears due to 
twisted sectors of the orientifold theory.
In ref. \khvost\ new branches of F-theory on K3 have been found which 
contain $T^4/{\bf Z}_3$, $T^4/{\bf Z}_4$ and $T^4/{\bf Z}_6$.
These special points correpond to $E_6\times E_6\times E_6$, 
$E_7\times E_7\times SO(8)$ and $E_8\times E_6\times SO(8)$ gauge groups 
respectively. 
A rigorous analysis of BPS states responsible for these enhanced gauge 
symmetries is however very complicated because of 
the essentially 
non-perturbative character of the $D_n$ and $E_n$
gauge symmetries.
More precisely in order to recover $E_n$ groups in terms of generalized 
orientifolds 
it is necessary to take into account the
``twisted sector'' associated with 
the element of the orientifold group involving $S$-duality 
transformation \khvost .

In the present paper we study 
the structure of BPS states responsible for enhanced gauge symmetries in the worldvolume of 7-branes.
While $A_n$ gauge groups have been 
already well understood \ref\witten{E. Witten,
Nucl. Phys. {\bf B 460} (1996) 335.}, \BSVa\
in terms of open strings connecting Dirichlet branes
one may wish to have a better insight into 
an appearance of $D_n$ and $E_n$ groups.
We formulate the local conditions for stable BPS states
in the worldvolume of 7-branes and identify 
those for $A_n$, $D_n$ and $E_n$ gauge groups 
with certain non-trivial $(p,q)$ 
open strings connecting appropriate 7-branes.
In general the relevant BPS states are found to correspond to 
strings connecting various 7-branes along rather complicated trajectories.
Since the open strings connecting the 7-branes are, in general, 
non-perturbative objects our analysis is
to some extent phenomenological.
Yet it is nice to see how the simple physical arguments
fit with the general picture of F-theory
compactifications.

{\it 2. Conditions for stable BPS states}

Consider a compactification of $F$ theory on an elliptic K3 \vafa .
The fibration of such a K3 can be described by a Weierstrass model 
\eqn\Wm{y^2=x^3+f(z) x+
g(z)=(x-e_1(z))(x-e_2(z))(x-e_3(z)),}
with the discriminant \ref\book{N. Koblitz,
{\it Introduction to Elliptic Curves and Modular Forms}.
Springer-Verlag, 1993.}
$\Delta=4f^3+27g^2=(e_1-e_2)^2(e_2-e_3)^2(e_3-e_1)^2.$
Here $z$ is a coordinate on $P^1$,
and $f$ and $g$ are polynomials of $z$ of degrees 8 and 12 respectively.
The 24  points on $P^1$ 
where the fiber ($T^2$) degenerates, i.e. the discriminannt $\Delta=0$
can be interpreted as 7-branes on
$P^1$.
Such a single 7-brane can be thought of as a 
(SL$(2,{\bf Z})$ transformed)
D-brane.
The positions of the 7-branes are mapped into the parameters of Wilson 
lines in the dual
heterotic description \vafa .
The 7-branes are the magnetic sources of 
a non-trivial classical background for 
$\tau=a+ie^{-\phi},$
where $a$ stands for an axion and $\phi$ is a dilaton.
The modular parameter $\tau$ is defined by $j(\tau)=4(24f)^3/\Delta.$

When $n$ Dirichlet 7-branes come to the same point on $P^1$ 
one gets an $SU(n)$ non-abelian gauge symmetry due to open strings 
connecting these 7-branes \witten .
In the F-theory 7-branes are generically mutually non-perturbative objects
\vafa .
As it has been shown in ref. \ref\mdli{Michael R. 
Douglas and Miao Li, `D-Brane Realization of N=2 Super Yang-Mills Theory in 
Four Dimensions', hep-th/9604041.}
from an analysis of
the low energy anomalous equations of motion
for $B_{NS,R}$ fields in the presence of 7-branes
branes a given 7-brane admits 
open strings with a unique $(p,q)$ charges
to end on it.
For example a Dirichlet 7-brane admits only $(1,0)$
(NS) strings to end on it.

In general the condition of ref.\mdli\
restricts possible enhanced gauge symmetries for a given configuration of 7-branes.
For the purpose of this paper
it is convenient to reformulate this condition as
follows:
a monodromy matrix which transforms the periods of a
$T^2$ fiber around the 
positions of the 7-branes has to leave the $(p,q)$ charges of the open string invariant.
Thus we essentially reduce the problem to an analysis 
of the
monodromies of the modular parameter $\tau$.  
In particular for the case of
an $A_{n-1}$ singularity \MV , \KMP\ 
it is easy to check that
the monodromies around $n$ 7-branes
commute with each other.

A very important subtlety is however that 
a definition of
monodromies for a given configuration of 7-branes 
implies a choice of a base point on a cover of $P^1$.
Therefore the same 7-brane on different sheets may be viewed as having different $(p,q)$ charges with respect to a chosen base point.
Hence two 7-branes of different charges may be connected 
by an open string through a complicated path.
This observation\foot{A relevance of such string configurations has been independently noted by 
Sen \ref\asen{A. Sen, `BPS States on a 
Three Brane Probe', 
hep-th/9608005.}.} is crucial for an explanation 
of an appearance of $E_n$ gauge groups (see below).

In order to understand this point better let us recall that from the analysis of singularities of 
type II theory 
on K3 an enhanced gauge symmetry 
appears due to solitonic states corresponding
to vanishing holomorphic 2-cycles of K3.
For the $A_n,$ $D_n$, $E_n$ singularities 
(resolved by an ALE space)
the cycles which shrink to zero area have 
intersection matrix corresponding to the Dynkin 
diagram for $A_n$, $D_n$, $E_n$
\BSVa , \KMP , \MV .
For an elliptically fibered K3 such a 
vanishing
2-cycle is locally a combination of 
an appropriate 1-cycle $b$ 
of the $T^2$ fibre and a 1-cycle on the $P^1$ base.
Its projection on $P^1$ is just an 
open string connecting appropriate 
branes.

We expect that the BPS states in F-theory on K3 are also 
associated with
such vanishing 2-cycles even though it may be 
considered as type IIB theory on $P^1$ \vafa ,\MV .
A projection of a 2-cycle to $P^1$ is represented by 
an open string connecting two 7-branes. 
Therefore when moving along a
contour $\gamma$ encircling such an open string
together with the two 7-branes (one in a clockwise direction,
another in an anti-clockwise direction)
the 1-cycle in the fiber $T^2$ has to go to itself.
That means that the total monodromy along $\gamma$ 
has to be just 1, i.e. $M^{-1}L^{-1} M' L=1$ where $M, M'$ are the monodromies around 
two 7-branes and $L$ stands for a monodromy along the corresponding open string.

Another condition comes out from the exisitence of a string of a minimal 
length connecting appropriate 7-branes.
The tension of a $(p,q)$ string reads
\john\
$T_{p,q}={1\over \sqrt{\tau_2}} |\tau_{p,q} (z)|,$
where $\tau_2={\rm Im} \tau$ and
$\tau_{p,q}=p+q\tau$.
The mass of a $(p,q)$ string state along a curve $C$ is given by
\asen
$$\int_C T_{p,q} ds= \int_C |dw_{p,q}|,$$
where the interval element $ds$ is given by \ref\cosmic{B. Greene, A. 
Shapere, C. Vafa and S.-T. Yau, Nucl. Phys. {\bf B 337} (1990) 1.}
$ds^2= \tau_2 |\eta (\tau)|^4 \prod_i |z-z_i|^{-1/6} |dz|^2,$
and 
$$dw_{p,q}= \eta(\tau)^2 \prod_i (z-z_i)^{-1/12} (p+ q\tau) dz.$$
Here $z_i$ stand for positions of 7-branes on $P^1$.
The 1-differential $dw_{p,q}$ can be represented as
$dw_{p,q} =\int_b \omega$, where $\omega$ is an appropriate closed
2-form on a ALE space.
An open string of minimal mass corresponds to
a straight line in the (flat) $|dw_{p,q}|^2$ metric.
Technically  
there should exist a geodesic
subject to 
$\int ^z dw_{p,q}=\alpha t$ which 
connects the two 7-branes, where $t$ is a proper time 
parameter along the open string and $\alpha$ is a constant.
This argument is similar to that
of ref. \ref\SD{A. Klemm, W. Lerche, P. Mayr,
C. Vafa and N. Warner, {\it Self-Dual Strings and N=2 Supersymmetric Field Theory}, hep-th/9604034.}.

In the simplest example of $n$ Dirichlet 7-branes
this condition is trivially satisfied because 
$dw\sim dz$.
It is also worth noting that in the presence 
of a non-Dirichlet 7-brane there is only one geodesic
that connects two Dirichlet 7-branes (see, 
for example Fig.1a
for a result of a numerical computation) while
the cases of $D_n$ and $E_n$ are more complicated
(see below).
In what follows we use the above observations 
to indentify the BPS states corresponding to the
$D_n$ and $E_n$ groups.

{\it 3. $D_n$ case}

The $D_4$ case has been studied in ref. \sen \
where it has been mapped to a $T^4/{\bf Z}_2$
orientifold.
In this subsection we briefly review this example 
and consider the general case of $D_n$ groups. 

In the $D_4$ case $f$ and $g$ are polynomials of 
degree 2 and 3 respectively \MV .
The discriminant 
$\Delta$ is a polynomial of order 6.
Hence there are six 7-branes.
The $SO(8)$ gauge group appears when $f\sim z^2$ and $g\sim z^3.$
To resolve the singularity one can take $f=z^2-{6\over q}z+{2\over 3} q,~~
g=2z^3+2qz+2,$ with $q^3=-27/2$, so that 
$\Delta \sim z^4 (5z^2-6z-1).$
This resolution corresponds to an unbroken $SU(4)$ subgroup of $SO(8)$ due to 
four 7-branes at $z=0$ 
(the type $I_4$ singularity \MV ).

The monodromy at infinity is given by $M_{\infty}=-1$.
We have $A^4=S^{-1}T^4S$ for 
a monodromy around $z=0$,  
$B=A^{-3}TA^3$ and $C=A^{-1}TA$ for monodromies 
around the roots of
the polynomial $5z^2-6z-1$.
One can easily check that $-1=A^4BC$.
The charges of the open strings corresponding to the
7-branes with $A$, $B$ and $C$ monodromies are
(0,1), (1,3) and (1,1) respectively.

The maximal $A_n$-type subgroup is $SU(4)$ which is due to the 4 mutually
perturbative 7-branes of the 
same type $A$ at $z=0.$
The monodromy around any of these four 7-branes 
does not commute with those around each 
7-branes of types $B$ and $C$ separately.
However it commutes with the monodromy along
a path encircling both  
$B$- and one of the $C$-type 7-branes.
The Chan-Paton factors for the open strings
stretched along such a non-trivial path are 
antisymmetric with respect to these 7-branes.
In particular this is clear in the 
$T^4/{\bf Z}_2$ orbifold limit, see 
\sen .
In the F-theory we observe that because 
of $BC=-A^{-4}$ the charges
change from
$(0,1)$ to $(0,-1)$ when one moves around
$B$ and $C$ 7-branes.
On the other hand the signs of the charges are related
to the gauge
charges of the ends of the open string on the 7-branes and, hence
to the orientation of the string \ref\green{Michael B. Green, Michael Gutperle, Phys. Lett.
{\bf B 377} (1996) 28, hep-th/9602077.},
\mdli .
Therefore locally at each of the 7-branes the string
has effectively different orientations.
This leads us to an identification of states
similar to
the case of a IIB theory on a $T^2$ orientifold.
As a consequence a natural 
interpretation of the $SO(8)$ gauge group is that 
its $SU(4)\times U(1)$ subgroup 
gets completed into the $SO(8)$ taking into 
account of states corresponding
to open strings which connect 
the perturbative 7-branes along the path that
encircles two non-perturbative $BC$ 7-branes \sen .
We then get a ${\bf 6}+\bar{\bf 6}$ 
representation (for the open strings with 
both orientations) of $SU(4)$ 
which is to complete the adjoint
representation of the $SO(8)$ 
gauge group.
Equivalently ${\bf 28}= {\bf 15} +{\bf 6} (-2) 
+{\bf \bar{6}}(+2)+{\bf 1}$,
where ($\pm 2$) are the charges with respect to the $U(1)$ gauge 
group on the two separated 7-branes.

It is now easy to construct a configuration of 7-branes which realizes an
$SO(2n+8)$ group (type $I^*_n$ singularity \MV).
Such a configuration consists of 
$n+4$ 7-branes with monodromy $A$, one 7-brane with monodromy $B$ and another one of monodromy $C$.
In this case the monodromy at infinity is $-A^n$ 
so that
\eqn\Dn{-A^n=A^{n+4}BC.}
The perturbative subgroup is in this case
$SU(n+4)\times U(1)$.
The adjoint representation of $SO(2n+8)$ 
reads as follows in terms of the above subgroup 
(we omit the $U(1)$ quantum numbers): 
${\bf (n+4)(2n+7)}= ({\bf (n+4)^2-1})+({\bf 1})+
({\bf {(n+4)(n+3)\over 2}}) + (\overline{\bf {(n+4)(n+3)\over 2}}).$
The non-perturbative states 
$({\bf {(n+4)(n+3)\over 2}}) + (\overline{\bf 
{(n+4)(n+3)\over 2}})$ are produced by strings 
connecting two
different $A$ 7-branes along a path that
encircles $BC$ 7-branes.
Because the matrix $BC$ acts with -1 on the 
$(p,q)$ charges of the open string 
and moreover we apply similar argument with 
the $D_4$ case with regard to the Chan-Paton indices
the above mentioned states belong
to the antisymmetric 2-tensor
representation of $SU(n)$.
Numerical work with {\it Mathematica} confirms
the existence of the corresponding geodesics.

In particular the $SO(32)$ group would be realized with 16=12+4 $A$ 7-branes 
in the presence of $B$ and $C$ 7-branes.
Note however that 
in terms of elliptic fibration of K3 
the configuration \Dn\
should correspond to $n=3{\rm deg}~f -
{\rm deg}~\Delta$, $3{\rm deg}~f=2{\rm deg}~ g.$
Note that for $n>4$ such a configuration of 7-branes cannot be realized
by a compactification of F-theory on K3 because the corresponding 
singularity destroys the triviality of the canonical bundle on a resolution \MV .

{\it 4. $E_n$ case}

We first consider a $E_6$ gauge group.
In this case we have 
(type IV$^*$ singularity \MV )
five 7-branes with the monodromy $A$, one 7-brane
with the monodromy $B$ and two 7-branes with monodromies
$C$, so that 
$$ST=A^5 BC^2,$$
where
$A=S^{-1}TS,~~~B=A^{-3}TA^3,~~~
C=A^{-1}TA.$
In the orientifold limit  
(F-theory on $T^4/{\bf Z}_6$)
this case was studied in ref. \khvost \
where it has been noted that in order
to have an $E_6$ group one needs to take into account
non-perturbative twisted sectors.
In what follows 
we identify the corresponding BPS states.

The perturbative subgroup is just $SU(5)\times SU(2)$.
Let us consider the non-perturbative states.
We observe that $BC$ commutes with $A$ and $A^4B$ commutes with $C$.
The matrices $BC$ and $A^4 B$ act with -1 on
the charges of the open strings corresponding 
to the $A$-type
and $C$-type 7-branes respectively.
Therefore one can consider 
an open string that connects an $A$ 7-brane 
with another one along a non-trivial 
path that encircles $B$ and one of the $C$ 7-branes.
This implies a representation $({\bf 10},{\bf 2})+(\overline{{\bf 10}},{\bf 2})$
according to the same argument as in the $D_n$ case.
These states are doublets of $SU(2)$ because one has to label the string according to which ones of the 
$C$ 7-branes is encircled:
the Weyl group of $SU(2)$ exchanges 
the $C$-type 7-branes.
Note that this
is the only possibility to construct a state 
with more than 2 Chan-Paton indices.
One can also 
consider an open string connecting the two different 
$C$ 7-branes along a non-trivial path encircling
$B$ and four of $A$ 7-branes.
This gives a representation $({\bf 5},{\bf 1})+
(\bar{{\bf 5}},{\bf 1}).$
Again the non-trivial quantum numbers with respect to $SU(5)$ 
correspond to a choice of four of $A$ 7-branes to be encircled.
It is easy to check by a numerical computation
that the corresponding geodesic exists. 
Thus together with the adjoint representation of the 
perturbative $SU(5)\times SU(2)\times 
U(1)$ gauge group we get an adjoint representation of $E_6$ group:
${\bf 78}=({\bf 24},{\bf 1})+({\bf 1},{\bf 3})+({\bf 1},{\bf 1}) +({\bf 10},{\bf 
2})+(\overline{{\bf
10}},{\bf 2})+({\bf 5},{\bf 1})+
(\bar{{\bf 5}},{\bf 1}).$

It is interesting 
to note that the maximal subgroup of 
$E_6$ is $SU(6)\times SU(2)$.
Therefore with the above configuration one 
realizes the $SU(6)$ 
subgroup non-perturbatively.
More precisely the $SU(6)$ adjoint representation consists of
the adjoint representation $({\bf 24},{\bf 1})+
({\bf 1},{\bf 3})+({\bf 1},{\bf 1})$
of the $SU(5)\times U(1)$ group
and the $SU(2)$ singlet 
BPS states $({\bf 5},{\bf 1})+(\bar{{\bf 5}},{\bf 1})$
which are 
produced by the open strings connecting two different 
$C$ 7-branes!
The higgsing of $E_6$ to $SO(10)\times U(1)$ corresponds to
a higgsing of the $SU(5)$ perturbative subgroup
to $SU(4)\times U(1)$.
Indeed, the unbroken perturbative subgroup gives now $({\bf 
15},{\bf 1})+ ({\bf 1},{\bf 3})+({\bf 1},{\bf 1})$.
The $AA$ strings encircling $BC$ give $({\bf 6},{\bf 2})+
(\bar{\bf 6},{\bf 2})$, while
the $CC$ strings encircling $A^4B$ produce 
$2({\bf 1},{\bf 1})$.
All together these representations are perfectly 
combined into the adjoint 
representation of $SO(10)$ which is thus non-perturbative.
This realization of $SO(10)$ is quite different from that of the previous section.
By higgsing one of $C$-type 7-branes out,
the
$({\bf 5},{\bf 1})+(\bar{{\bf 5}},{\bf 1})$ states
and half 
of the $({\bf 10},{\bf 2})+(\overline{{\bf 
10}},{\bf 2})$ ones become heavy and
we recover a realization of a
${\bf 45}={\bf 24}+{\bf 10}+\overline{\bf 10}+{\bf 1}$ representation of 
$SO(10)$ discussed in the previous section.

The structure of the BPS states for $E_7$ and $E_8$ gauge groups is more complicated.
Consider six 7-branes with the monodromy $A$, 
one 7-brane
with the monodromy $B$ and two 7-branes with 
monodromies $C$, so that
$$S=A^6 BC^2 .$$ 
In the orientifold limit\kucha \ this case 
corresponds to $T^4/{\bf Z}_4$
and has to realize a $E_7$ group 
(type III$^*$ singularity).

The perturbative gauge group is just $SU(6)\times SU(2)\times U(1)$ with the adjoint representation
$({\bf 35},{\bf 1})+({\bf 1},{\bf 3})+ 
({\bf 1},{\bf 1})$.
Similar to the above construction we can get additional states 
$({\bf 15}, {\bf 2})+ (\overline{\bf 15},{\bf 2})$ 
due to 
open strings connecting $A$ 7-branes 
along a non-trivial path 
which encircles $B$ and one of the $C$ 
7-branes.
For strings connecting the $C$-type 7-branes along
a path encircling $B$ and four of the
$A$ 7-branes one gets
$({\bf 15},{\bf 1})+(\overline{\bf 15},{\bf 1}).$
The adjoint representation of $SU(6)\times U(1)$ 
subgroup together with the
$({\bf 15},{\bf 1})+(\overline{\bf 15},{\bf 1})$ 
one gives the adjoint representation of
the non-perturbative $SO(12)$.
However all of
the 129 states listed above do not complete 
the adjoint representation ${\bf 133}$ of $E_7$.
The missed states are $2({\bf 1},{\bf 2}).$
Together with $({\bf 15}, {\bf 2})+ 
(\overline{\bf 15},{\bf 2})$ these are
combined into $({\bf 32}',{\bf 2})$ of 
$SO(12)\times SU(2).$ 

As it has been anticipated above
the additional states can come from 
string connecting $B$ and $C$ 7-branes along a  non-trivial 
way so that the open string meets these 7-branes on different sheets.
The point is that a $B$ 7-brane corresponds to the $(p,q)$ charges of a $C$ 
7-brane on another sheet.
Numerical computations with {\it Mathematica}
confirm the existence of the corresponding geodesic
schematically shown in Fig.1b.

\vskip .2in
\let\picnaturalsize=N
\def\picsize{4.0in}
\def\picfilename{fig1.eps}
\ifx\nopictures Y\else{\ifx\epsfloaded Y\else \fi
\global\let\epsfloaded=Y
\centerline{\ifx\picnaturalsize N\epsfxsize \picsize\fi
\epsfbox{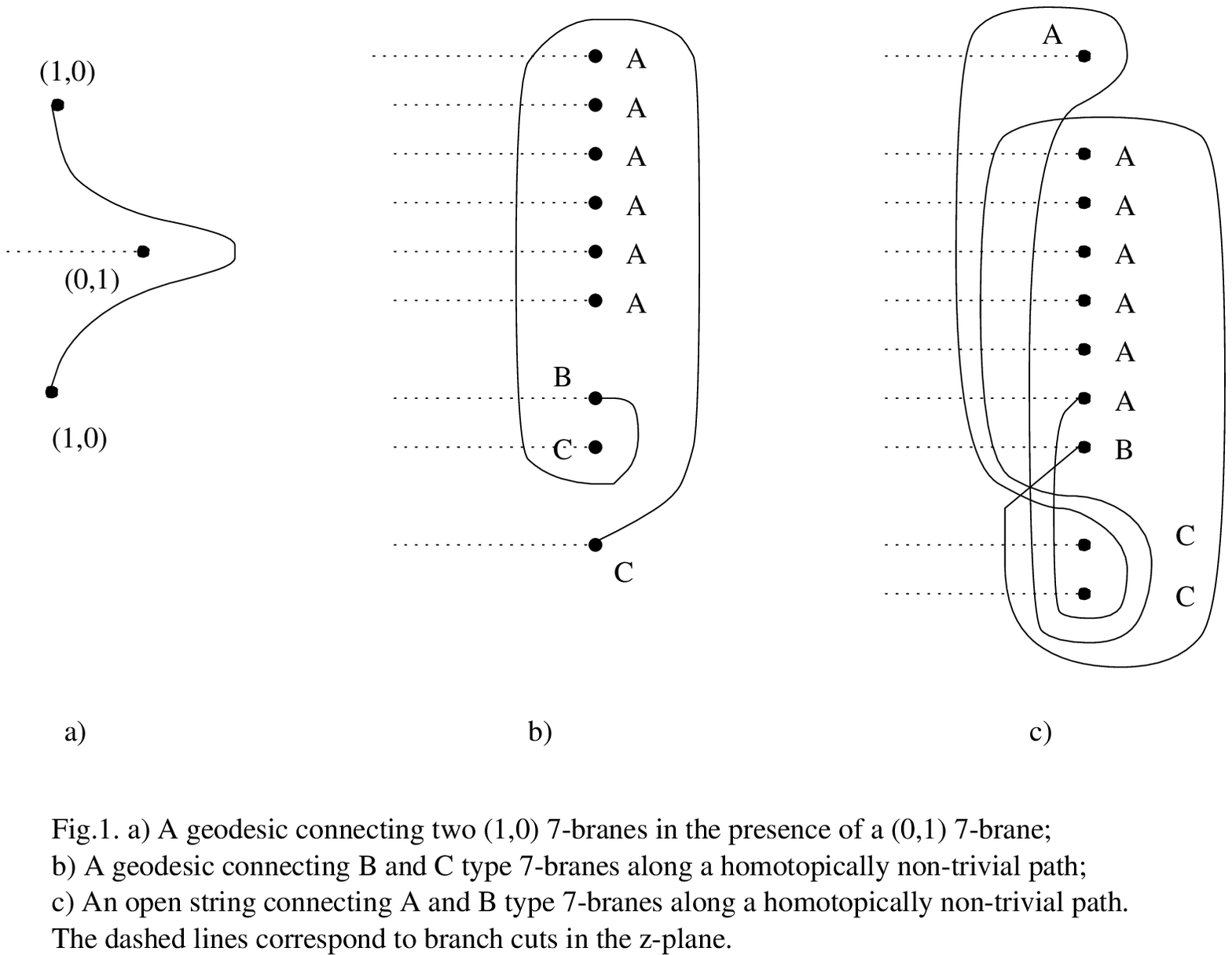}}}\fi

It corresponds to $C^{-1}L^{-1}BL=1$ where 
$L=CS,$
so that this matrix transform the $(1,1)$ charges of the $C$ 7-brane 
into the
$(1,3)$ charges of the $B$ 7-brane.
These states transform as $2({\bf 1},{\bf 2})$.
We thus get the ${\bf 133}$ representation of $E_7$.
Note that for such a string we need 6 $A$-type 7-branes.
Therefore such an additional state was absent in the $A^5BC^2$ case.
When one of the $A$-type 7-branes is higgsed 
out these additional states become heavy and
we recover
an embedding $E_7\to E_6\times U(1)$.

Consider now the configuration corresponding to 
a $E_8$ group (type II$^*$ singularity) which 
consists of
seven 7-branes with the monodromy $A$, one 7-brane
with the monodromy $B$ and two 7-branes with 
monodromies $C$, obeying
$$ST=A^7 BC^2 .$$
In the orbifold limit this configuration corresponds to 
$T^4/{\bf Z}_6$ \kucha \ as follows from $(ST)^6=1$. 
The perturbative subgroup is just $SU(7)\times SU(2)\times U(1)$
with the gauge bosons in the $({\bf 48},{\bf 1},{\bf 0})+({\bf 1},{\bf 3},{\bf 0})+({\bf 1},{\bf 1},{\bf 0})$
representation.
The $AA$ strings encircling the $BC$ 
7-branes
give $({\bf 21},{\bf 2})+(\overline{\bf 21},{\bf 2})$ 
representation of $SU(7)\times SU(2)$.
The $CC$ strings encircling $A^4B$ give $({\bf 35},{\bf 1})+(\overline{\bf 
35},{\bf 1}).$
The adjoint representation ${\bf 248}$ of $E_8$ is expanded in terms of 
representations of the perturbative group $SU(7)\times SU(2)$ as follows
$${\bf 248}=({\bf 48},{\bf 1})+({\bf 1},{\bf 3})+({\bf 1},{\bf 1})+({\bf 21},{\bf 2})+(\overline{\bf 
21},{\bf 2})
+({\bf 35},{\bf 1})+(\overline{\bf
35},{\bf 1})+$$
$$+({\bf 7},{\bf 1})+(\bar{\bf 7},{\bf 1})+ ({\bf 7},{\bf 2})+
(\bar{\bf 7},{\bf 2}).$$
The missed representation is thus $({\bf 7},1)+(\bar{\bf 7},{\bf 1})+ ({\bf 
7},{\bf 2})+
(\bar{\bf 7},{\bf 2}).$
The representation $({\bf
7},{\bf 2})+
(\bar{\bf 7},{\bf 2})$ is due to 
the string connecting $B$ and $C$ 7-branes along a path that encircles 
six of seven 7-branes similar to that
shown in Fig.1b.
The representation $({\bf 7},{\bf 1})+(\bar{\bf 7},{\bf 1})$ seems to correspond to 
a string connecting $A$ and $B$ 7-branes
along a complicated path shown in Fig.1c.
This path corresponds to $B=L^{-1}AL$ where 
$L=S^{-1}C^2A^{-1}S.$
This matrix transform the $(1,3)$ charges of the $B$ 7-brane into 
$(0,1)$ charges of the $A$ 7-brane.
Assuming that this representation is correct one
can move one of $A$-type 7-branes out.
Then the BPS state $({\bf 7},{\bf 1})+(\bar{\bf 7},{\bf 1})$ becomes heavy and only $2({\bf 1},{\bf 2})$ 
part of $({\bf 7},{\bf 2})+
(\bar{\bf 7},{\bf 2})$ remains light.
Thus we recover an embedding $E_8\to E_7\times U(1)$.

{\it 5. Conclusion and Acknowledgements}

We have identified stable BPS states 
relevant for $D_n$ and $E_n$
enhanced gauge symmetries in the F-theory
in 8 dimensions with particular
open string configurations connecting appropriate
7-branes.
As a consistency check of the above picture
we have recovered various embeddings
$E_8\to E_7\times U(1)$, $E_7\to E_6\times U(1),$ $SO(12)\times U(1)$,
$SU(7)\times U(1)$, $E_6\to SO(10)\times U(1),$ $SU(6)\times U(1)$.

The simple physical arguments nicely
fit with results which follows from the analysis of
singularities and duality between F-theory on K3 and
heterotic string on $T^2$.
\vskip .1in

Finally I wish to thank M. Bershadsky, S. Sethi and
C. Vafa for useful conversations.
I also thank A. Krasnitz for useful comments on 
numerical computations.
I am grateful to the Niels Bohr Institute where 
this work has been done for hospitality.
This work is partially supported by the
Packard Foundation and by NSF grant PHY-92-18167
and by NATO grant GRG 93-0395.

\listrefs
\end